\documentstyle[frontieres,art10]{article} 
\def\lapp{\ifmmode\stackrel{<}{_{\sim}}\else$\stackrel{<}{_{\sim}}$\fi} 
\def\gapp{\ifmmode\stackrel{>}{_{\sim}}\else$\stackrel{>}{_{\sim}}$\fi} 
\begin{document} 
\heading{Pulsar Searches --- {\it Tricks of the Trade}}  
\par\medskip\noindent 

\author{Dunc Lorimer \\ \\
Max-Planck-Institut f\"ur Radioastronomie, Auf dem H\"ugel 69, D-53121, Bonn} 
 
\begin{abstract}
The study of pulsars has come a long way since their accidental discovery 
as ``scruff'' on pen chart recordings of the radio sky at 81 MHz by
Jocelyn Bell and Anthony Hewish at Cambridge some thirty years
ago. The present sample of almost 800 objects is the result of an
immense amount of observing time at the largest radio telescopes
around the world. This article highlights the most exciting finds, and
summarises the most commonly used, yet relatively simple, techniques
which continue to be most efficient at finding weak periodic signals
hidden in noisy time series.
\end{abstract} 

\section{Preamble}
\label{sec:preamble}

Following the initial discoveries at Cambridge \cite{hbp+68,phbc68,cp68},
a number of pulsars were unearthed by groups around the world in the
next few years. As in the Cambridge survey, many of these early
discoveries were made by visual inspection of pen chart recordings
collected from the detected output of a large radio telescope. By
definition, this method selects against those objects too faint to be
detected from their individual pulses. To find such sources, which are
likely to make up the bulk of the potentially observable sample, we
require a more sophisticated search algorithm which can search for
{\it a priori} unknown periodic signals in a longer time series.
Following some success using algorithms developed to fold data in the
time domain \cite{sta69}, it was realised that a more efficient method
is to take the Fast Fourier Transform (FFT) of the time series. A
periodic signal is then easily identified by a characteristic set of
harmonics in the power spectrum. An additional complication arises
when searching for more distant pulsars whose pulses become smeared
across a finite receiver bandwidth as a result of their passage
through the ionised component of the interstellar medium. As we shall
see (\S \ref{sec:search}) this can largely be removed by splitting the
pass-band up into a finite number of channels and applying appropriate
time delays to each channel before summing over the channels. Since
the magnitude of this effect is unknown {\it a priori}, a number of
time series are formed and FFTed. This basic two-dimensional approach,
which has been used almost exclusively in all searches since the early
1970s, will be described in detail in the following.

\section{A Potted History of Pulsar Searching}
\label{sec:hist}

\begin{figure}[hbt]
\setlength{\unitlength}{1in}
\begin{picture}(0,2.6)
\put(-0.2,3.05){\includegraphics{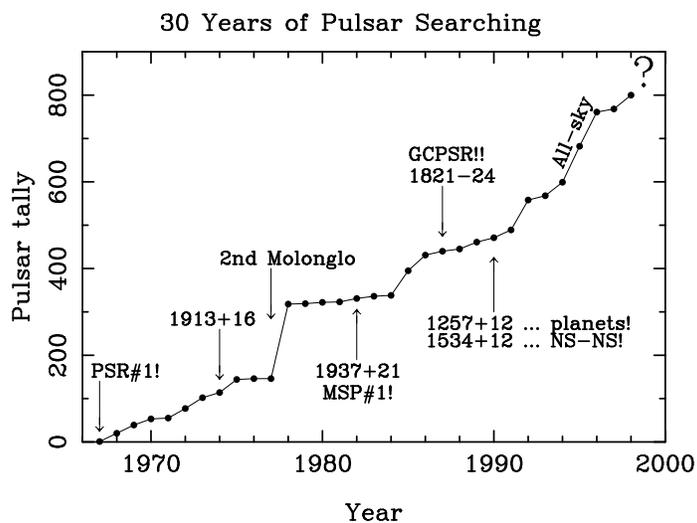}}
\end{picture}
\caption[]
{\sl
The cumulative number of pulsars known as a function of time. This
diagram captures some of the ``thrill of the chase'', as well as the
continued motivation to initiate new searches following significant
technological advances and conceptual innovations.  See text for
discussions of individual objects.
}
\label{fig:tally}
\end{figure}

One means of summarising the history of progress in pulsar searches is
the cumulative number distribution in Fig.~\ref{fig:tally}. This
shows, amongst other things, where landmark discoveries of exciting
pulsars have been made, and their impacts on subsequent searches.
Perhaps the most famous discovery of all of these is the original
binary pulsar, PSR B1913+16\footnote{Pulsars are conveniently named
with a PSR prefix followed by their celestial coordinates.},
discovered by Hulse \& Taylor in 1974 in a survey of part of the
galactic plane visible from the giant 305--m Arecibo radio telescope
\cite{ht75a,ht75b}. This relativistic binary system, which consists of
a pair of neutron stars, has been shown to be a truly remarkable
natural laboratory for tests of general relativity \cite{tw82,tw89}
--- certainly more than ample reward for the initial outlay of
telescope time in the survey and subsequent off-line processing.
Fascinating accounts of the discovery and study of this system can be
found in the 1993 Nobel lectures given by Hulse and Taylor \cite{hul94,tay94}.

In 1977 Dick Manchester and colleagues used the Molonglo radio
telescope in Canberra to survey the sky with declinations $-85^{\circ}
\leq \delta \leq +20^{\circ}$ ($\sim 8.4$ sr).  This monumental effort
\cite{mlt+78} discovered no less than 155 pulsars, thereby doubling the
observed pulsar population known at that time (the 1977/8 leap in
Fig.~\ref{fig:tally}). The new discoveries included one more binary
system PSR B0820+02 --- an ordinary 0.865 s pulsar in a wide orbit
around a white dwarf star \cite{mncl80}.  Together with the results of
contemporary searches using northern hemisphere telescopes
\cite{dls77,dth78}, a fairly well-understood sample of pulsars over
the whole sky was available for studying the statistical properties of
the Galactic population \cite{pb81,vn81,lmt85}. These studies suggested
a total population of $\sim 10^5$ active radio pulsars in the Galaxy.

Data acquisition and processing limitations during the 1970s and early
1980s limited the sampling rates of surveys conducted during this era
to $\gapp 20$ ms. The corresponding sensitivity to short-period
objects, such as the 33 ms pulsar in the Crab nebula, was thus far
from ideal suggesting that the true population of short-period pulsars
was being underestimated in the observed sample. This was dramatically
confirmed in 1982 with the discovery by Backer et al.~of ``the
millisecond pulsar'' B1937+21 \cite{bkh+82,bac84}.  With a period of
1.56 ms and a corresponding rotation frequency over 20 times larger
than the Crab, this remarkable object is still the most rapidly
rotating neutron star known to man. Subsequent timing observations of
B1937+21 soon showed that it is an extremely stable celestial clock on
time-scales $\sim$ years \cite{dtwb85} having a host of astrophysical
applications including the detection of long-period gravitational
waves from the early Universe \cite{bcr83}.  In addition, PSR B1937+21
is a bright source, having a mean luminosity roughly 240 times that of
the original Cambridge pulsar, B1919+21. It thus seemed natural to
suppose that B1937+21 was just the tip of the iceberg of a larger
population of rapidly rotating neutron stars missed by previous searches.

The large increase in computing power and data storage requirements in
a search for millisecond pulsars meant that most early search efforts
had only limited sensitivity to millisecond pulsars. This problem is
highlighted by the fact that only 4 millisecond pulsars were found in
the galactic disk prior to 1990 which, in turn, hampered early attempts
to determine their galactic population \cite{kn88}. Surveys conducted
during this period were, however, very successful at discovering young
pulsars along the galactic plane \cite{clj+92,jlm+92} (see \S \ref{sec:sens}).

Following initially unsuccessful searches of globular clusters prior
to the discovery of PSR B1937+21 \cite{sg80}, search systems armed
with faster sampling rates returned to globular clusters --- where
low-mass X-ray binaries, the probable progenitors of millisecond
pulsars \cite{acrs82}, were already known to exist \cite{cla75}. This
approach circumvents the need to cover large areas of sky, since each
cluster can be observed with one telescope pointing, thereby greatly
reducing the total amount of data to process.  Searches soon proved
fruitful, with the discovery of PSR B1821--24, a 3.1 ms pulsar in the
globular cluster M28 \cite{lbm+87}. Surveys of other clusters have
since been very successful, discovering over 20 millisecond
pulsars. Notable highlights were the discovery of an eclipsing binary
system in Terzan 5 \cite{lmd+90}, 11 millisecond pulsars in
47--Tucanae \cite{mlr+91,rlm+95}, and a neutron star--neutron star
binary in M15 \cite{pakw91,and92}.

Advances in low-cost computing power and data storage capabilities
towards the end of the 1980s meant that a return to galactic disk
surveys with much improved sensitivity was possible. The breakthrough
was made by Wolszczan in a search of just 200 deg$^2$ of sky away from
the galactic plane during an upgrade period at Arecibo
\cite{wol91a}. The survey found the millisecond pulsar planetary
system B1257+12 \cite{wf92,wol94}, and yet another neutron
star--neutron star binary system PSR B1534+12 \cite{wol91a}. A
statistical analysis by Johnston \& Bailes \cite{jb91} demonstrated
that a large number of millisecond pulsars would be found by an
all-sky search of similar sensitivity to Wolszczan's survey.

\begin{figure}[hbt]
\setlength{\unitlength}{1in}
\begin{picture}(2,1.2)
\put(-0.6,1.7){\includegraphics{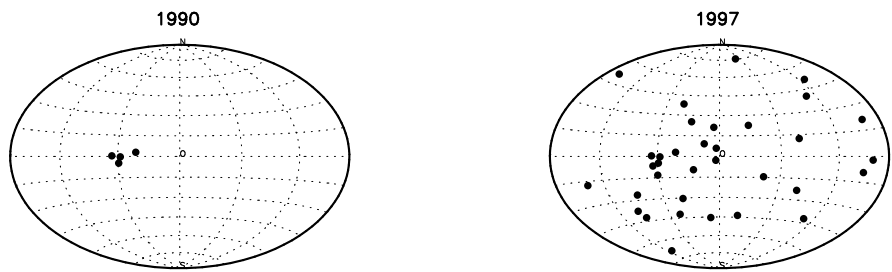}}
\end{picture}
\caption[] 
{\sl 
Hammer-Aitoff projections showing the known Galactic disk millisecond
pulsar population in 1990 (left) and in 1997 (right).
}
\label{fig:mspait}
\end{figure}

Large-area searches began in earnest in the early 1990s at Parkes
\cite{mld+96,lml+98}, Jodrell Bank \cite{nll+95}, Arecibo
\cite{tdk+93,fcwa95,rdk+95,cnt96,cnst96}, and Green Bank \cite{snt97}.
With the exception of the Arecibo surveys, these have now been
completed. The great success of these surveys can be seen by the sharp
rise in Fig.~\ref{fig:tally} during the mid 1990s, and also in
Fig.~\ref{fig:mspait} which compares the sky distributions circa 1990
with the present situation in which the sources are much more
uniformly distributed on the sky.  Over 30 millisecond pulsars have so
far been discovered by these searches as well as many more
low-luminosity ordinary pulsars. Notable highlights include PSRs
J0437--4715 \cite{jlh+93} and J1713+0747 \cite{fwc93} --- bright,
nearby millisecond pulsars which are already proving to be more stable
clocks than PSR B1937+21.  A future application of such an ``array''
of clocks will be as a very sensitive detector of long-period
gravitational waves \cite{fb90}.

Although no further pulsar planetary systems like B1257+12, or
relativistic binary systems like B1534+12, have thus far been found,
many of the newly discovered millisecond pulsars have white dwarf
companion stars which can be studied optically \cite{vbk96,lcf+96}.

\section{Practical considerations in pulsar searches}
\label{sec:search}

Having gotten a flavour for the potential rewards gained in finding
new pulsars, let us turn our attention to the specific problems
involved in the search. As mentioned in \S \ref{sec:preamble}, for any
given point on the sky, our two basic unknown parameters are the pulse
period and the amount of dispersion across the receiver bandwidth.  A
powerful technique to search for unknown periodicities, that is now
the ``industry standard'', is to take the Fast Fourier Transform
(FFT) of the noisy time series.  For any periodic signal with a small
duty cycle (5--10\% is typical of most pulsars), the resulting
amplitude spectrum from the FFT consists of a family of harmonic
spikes with the fundamental corresponding to the signal
frequency. Having identified the fundamental, the original time series
can be folded at the apparent pulse period to form an integrated
profile with a high signal-to-noise ratio (SNR).

By considering the train of weak pulses in the time series to be a set
of equally-spaced delta functions convolved with a simple boxcar
function as the pulse profile, the number of harmonics in the power
spectrum of the FFT can be shown to be equal to the inverse of the
pulse duty cycle.  Thus, much of the total power of a narrow pulse is
distributed in its higher order harmonics, rather than just the
fundamental. In this case, we can improve greatly on the sensitivity
by adding higher harmonics onto the fundamental. For example, to add
all 2$^{\rm nd}$ harmonics onto their corresponding fundamentals, we
stretch the lower half of the amplitude spectrum by a factor of two
and add this to the original unstretched spectrum.
\begin{figure}[hbt]
\setlength{\unitlength}{1in}
\begin{picture}(2,2.7)
\put(-0.15,-3.7){\includegraphics{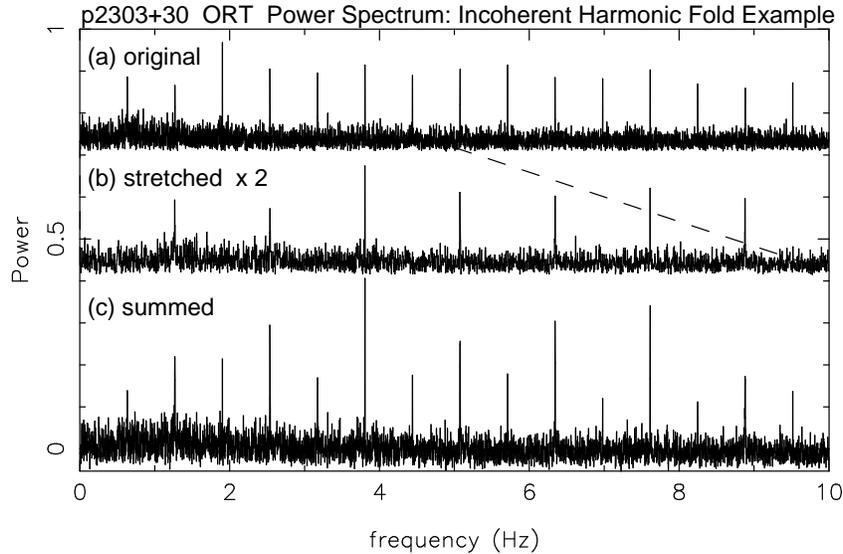}}
\end{picture}
\caption
{\sl
The process of harmonic summing illustrated in the power spectrum on
test data collected for PSR B2303+30, collected with the Ooty radio
telescope at 327 MHz. Figure kindly provided by Dipankar Bhattacharya.
}
\label{fig:hsum}
\end{figure}
An example of such a ``harmonic summing'' process is shown graphically
in Fig.~\ref{fig:hsum}. Note that, whilst the summation process
naturally increases the noise in the folded spectrum by a factor of
$\sqrt{2}$, the amplitude of the signal may increase by a larger
factor thus giving a net increase in the observed SNR. The action of
repeating this process several times (in typical pulsar searches
usually three further operations are done) is to effect a search in
pulse duty cycle. Typically, the majority of weak signals with small
duty cycles are only apparent in the harmonically folded spectra.

The group velocity of the pulsed radiation through the ionised interstellar
medium is frequency dependent: pulses emitted at higher radio
frequencies travel faster and arrive earlier than those emitted at
lower frequencies. This dispersion process has the effect of
``stretching'' the pulse across a finite receiver bandwidth, thereby
reducing the SNR.  The delay $\Delta t$ in arrival times between a
high frequency $\nu_{\rm hi}$ (MHz) and a low one $\nu_{\rm lo}$
(MHz), can be shown to be
\begin{equation}
 \Delta t = 4.15 \times 10^6 \, \, {\rm ms} \, \,
 \times (\nu_{\rm lo}^{-2} - \nu_{\rm hi}^{-2})  
 \times {\rm DM},
\label{equ:defdm}
\end{equation}
where the dispersion measure DM (cm$^{-3}$ pc) is the integrated
column density of free electrons along the line of sight to the
pulsar. Pulsars at large distances have higher column densities and
therefore larger DMs than those pulsars closer to Earth so that, from
Eq.~\ref{equ:defdm}, the dispersive delay across the bandwidth is
greater. A measurement of the DM, together with a suitable model for
the Galactic distribution of free electrons, can provide estimates for
the distances to pulsars \cite{tc93}.  The DMs in the present sample
of pulsars range between 2.4 cm$^{-3}$ pc (roughly 100 pc distant) out
to 1074 cm$^{-3}$ pc (roughly the distance of the Galactic centre)
with a median value of 82 cm$^{-3}$ pc (roughly 2 kpc).

\begin{figure}[hbt]
\setlength{\unitlength}{1in}
\begin{picture}(2,1.6)
\put(-1.7,3.7){\includegraphics{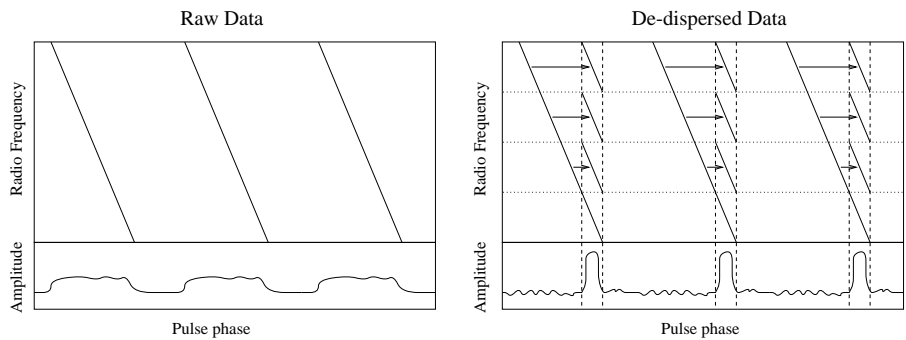}}
\end{picture}
\caption
{\sl
Pulse dispersion and the process of de-dispersion. The effect of
simply summing the pulse train over a finite bandwidth is to
significantly broaden the observed pulse (left panel). By dividing the
passband into smaller bandwidth channels and applying the appropriate
delay to each channel considerably reduces the broadening and
increases the pulse signal--to--noise ratio (right panel).
}
\label{fig:dedisp}
\end{figure}

In the extreme case, if we na\"{\i}vely neglect this effect in our
search algorithm, then there is a limiting DM above which the pulse
from a pulsar is stretched across the receiver bandwidth by an amount
greater than one period so that, regardless of flux density, such a
dispersed pulsar would be undetectable.  To see this quantitatively,
we can rearrange Eq.~\ref{equ:defdm} for a bandwidth $\Delta \nu =
\nu_{\rm hi} - \nu_{\rm lo}$ about a centre frequency $\nu =
\frac{1}{2}(\nu_{\rm lo} + \nu_{\rm hi})$ to obtain the ``dispersion
broadening'' relation (for $\nu \gg \Delta \nu$) as:
\begin{equation}
\label{equ:defsmear}
 t_{\rm DM} = 8.3 \times 10^6  \, {\rm DM} \, \frac{\Delta \nu}{\nu^3}
			     \, \, \, {\rm ms},
\end{equation}
where the centre frequency $\nu$ and bandwidth $\Delta \nu$ are
measured in MHz and the DM is in units of cm$^{-3}$ pc.  For example,
the first pulsar to be discovered, PSR B1919+21, has a relatively low
DM of 12.4 cm$^{-3}$ pc. Over the 1 MHz bandwidth of the original
Cambridge equipment observing at 81 MHz, the dispersion broadening is
$\sim$ 190 ms --- quite unacceptable for millisecond pulsar hunters!
Since this pulsar has a period of 1337 ms it was (fortunately)
detected. To compensate for this effect, modern searches divide the
full pass-band into a number of smaller bandwidth channels, over which
the dispersion broadening is correspondingly less (see
Fig.~\ref{fig:dedisp}).  By applying appropriate time lags to each
channel (usually in software), the time series can be de-dispersed for
large number of trial DM values. During the Parkes Southern Sky Survey
\cite{mld+96,lml+98}, for example, each time series was de-dispersed
for 738 DMs between 0 and 768 cm$^{-3}$ pc.

\begin{figure}[hbt]
\setlength{\unitlength}{1in}
\begin{picture}(0,1.9)
\put(-0.2,2.8){\includegraphics{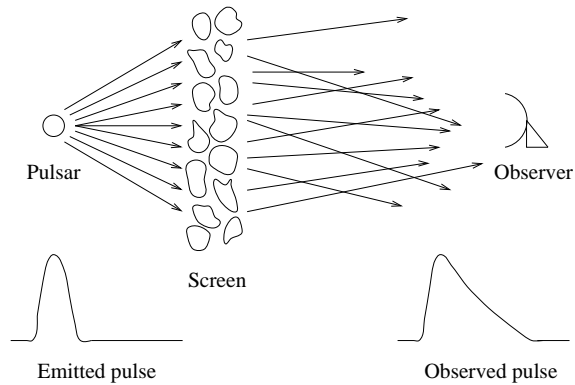}}
\end{picture}
\caption[]
{\sl
Diagram showing the ``thin--screen'' model of pulse scattering.
The difference in path lengths and
therefore in arrival times of the scattered rays result in a
``scattering tail'' in the observed pulse profile.
}
\label{fig:scatt2}
\end{figure}

As well as the dispersion broadening effect, free electrons in the
interstellar medium can scatter the pulses causing an additional
broadening due to the different arrival times of scattered pulses.  A
simple scattering model is shown in Fig. \ref{fig:scatt2} in which the
scattering electrons are assumed to lie in a thin--screen between the
pulsar and the observer \cite{sch68}.
At observing frequencies $\lapp$ 400 MHz, scattering becomes
particularly important for pulsars with DMs $\gapp$ 200 cm$^{-3}$ pc,
where the increased column density of free electrons can cause a
significant tail in the observed pulse profile as shown in
Fig.~\ref{fig:scatt2}, reducing the SNR.  The amount of scattering can
be shown to scale as $\nu^{-4}$. Thus, high frequency ($\gapp$ 1 GHz)
surveys are much less prone to scattering than $\sim$ 400 MHz searches.

To summarise our discussion on pulse broadening, we may consider the
observed pulse width $W$ as being the convolution of the intrinsic
width $W_{\rm int}$ emitted at the pulsar with additional broadening
functions due to dispersion, scattering, and by the post-detection
integration performed in the receiver. This can be expressed
approximately by the following quadrature sum:
\begin{equation}
\label{equ:weff}
  W^2 = W_{\rm int}^2 + t_{\rm samp}^2 + t_{\rm DM}^2 + t_{\rm scatt}^2,
\end{equation} 
where $t_{\rm samp}$ is the data sampling interval, $t_{\rm DM}$ is
the dispersion broadening across one filterbank channel and
$t_{\rm scatt}$ is the interstellar scatter broadening. 

\section{Search Sensitivity and Optimisation}
\label{sec:sens}

The limiting flux density\footnote{Observed 400 MHz flux densities vary
between 0.1 and 5000 mJy, the median is 12 mJy.}, $S_{\rm min}$,
required to detect a pulsar with a radio telescope can be written as:
\begin{equation}
\label{equ:smin}
   S_{\rm min} \simeq \frac{10 \, T_{\rm sys}}{G \sqrt{n \Delta
        \nu \tau}} \left(\frac{W}{P-W} \right)^{1/2}.
\end{equation}
Here $G$ is the forward gain of the antenna (K/Jy); $n$ is the
number of polarisations summed (usually 2); $\Delta \nu$ is the
observing bandwidth (MHz); $T_{\rm sys}$ is the system temperature (K)
--- essentially the sum of the thermal noise in the receiver (25--50 K
is typical for modern systems) and the excess noise from the radio
continuum background along the Galactic plane (see below); $\tau$ is
the integration time (s); $P$ is the period of the pulsar (s) and, as
defined in \S \ref{sec:search}, $W$ is the observed width of the pulse
(s). The additional factor of 10 is approximate, and takes into
account a limiting SNR required for a detection ($\sim 8$ for typical
pulsar searches), as well as losses in sensitivity due to hardware
limitations.  With this choice of units, $S_{\rm min}$ is in mJy (1
mJy $\equiv 10^{-29}$ W m$^{-2}$ Hz$^{-1}$).

To get a feel for the numbers, consider the Jodrell Bank 400 MHz
search system used in a recent northern sky survey for millisecond
pulsars \cite{nll+95}. Here $G = 1.0$ K/Jy, $n=2$, $\Delta \nu = 8$
MHz, $T_{\rm sys} \gapp 70$ K, which gives $S_{\rm min} \simeq 5$ mJy
for an integration time of 5 minutes.  Although the situation improves
as $\tau^{1/2}$ for longer integration times, a price is ultimately
paid through loss of sensitivity to pulsars in short-period ($\lapp$
few hr) binary systems (\S \ref{sec:binaries}), not to mention rapid
loss of popularity with the telescope's time allocation committee!

When planning a pulsar survey, we are faced with a number of
combinations of observing parameters which can significantly effect
the ultimate success/failure of the experiment. Before embarking on a
search, we should perhaps first ask ourselves: What sort of pulsars
are we interested in finding and where's the best place to look for
them? Broadly speaking, pulsar surveys to answer these questions fall
into one of three main categories:

\smallskip
--- A large area search for old and/or nearby low-luminosity pulsars.

--- A galactic plane search for young and/or distant luminous pulsars.

--- A targeted search towards supernova remnants, globular clusters {\it etc.}
\smallskip

\noindent
Let us consider the first two cases (the third category depends on the
location of the target objects with respect to the Galactic plane),
assuming that we have got observing time on a large radio telescope
equipped with a system that can sample fast enough to detect
short-period pulsars, and that we have settled on an integration time
that allows us to cover our search area within the time allocated to
us by the committee.

Initially, we may choose to survey exclusively at 400 MHz.  This does
not seem unreasonable, particularly since pulsar flux density spectra
are rather steep, with typical spectral indices of --1.6 \cite{lylg95}
and are known to turn over at frequencies below 400 MHz. For large
area surveys, this spectral consideration is the main factor since the
larger solid angle of the telescope beam at low frequencies allows a
faster rate of sky coverage than a single beam higher frequency
system.  In surveying the galactic plane at 400 MHz, however, we
immediately run into two big problems:

\smallskip
\noindent
{\it (i).} The system temperature, $T_{\rm sys}$, becomes dominated by
the sky background radiation: From the combined all-sky 408 MHz radio
continuum survey \cite{hssw82}, we infer extremely large values:
$\sim$ 900 K in the direction of the galactic centre, and $\sim$ 300 K
along the galactic plane.

\smallskip

\noindent
{\it (ii).} The observed pulse width $W$ in Eq.~\ref{equ:weff} becomes
dominated by the $t_{\rm DM}$ and/or $t_{\rm scatt}$ terms due to
excessive scattering and/or dispersion of pulses by free electrons in
the interstellar medium. In the extreme case $W \gapp P$, so that such
a pulsar would no longer be detectable by a time series analysis. For
dispersion, it may be possible, at a cost of additional processing
time, to increase the number of channels in the filterbank.
\smallskip

\noindent
Fortunately all these effects diminish strongly when we choose a
higher observing frequency: The spectral index of the radio continuum
emission is roughly --3, so that the sky background temperatures
quoted above are reduced by more than an order of magnitude for high
frequency ($\gapp 1$ GHz) surveys. As mentioned in \S
\ref{sec:search}, dispersion and scattering scale as $\Delta
\nu/\nu^{3}$ and $\nu^{-4}$ respectively thus becoming negligible when
moving to high frequencies.

Clifton et al.~\cite{clj+92} were the first pulsar hunters to really
exploit this approach, finding 40 new young pulsars in a 1400 MHz
survey of a thin strip of 200 deg$^2$ along the galactic plane.  All
of these pulsars were missed by a previous 390 MHz survey
\cite{stwd85} which overlapped the same region which had twice the
nominal sensitivity of the Clifton et al.~survey. Many of the pulsars
discovered in the high frequency survey are simply not observable at
frequencies much below 600 MHz due to a combination of the sky
background, dispersion and scattering effects mentioned above.  A
complementary survey of the southern Galactic plane using the Parkes
radio telescope at 1520 MHz \cite{jlm+92} also found 46 pulsars that
were missed by previous lower frequency searches covering this region
\cite{mlt+78}.

The main disadvantage of the high frequency surveys was always the
rate of sky coverage which scales with the inverse square of the
observing frequency. The current generation of 1400 MHz searches at
Parkes and Jodrell Bank are addressing this problem by installing
multi-beam receivers. At Parkes, a 13-beam system
can cover the sky at the same rate as the
recent Parkes 430 MHz survey \cite{mld+96,lml+98}. A 4-beam system is
presently being installed at Jodrell Bank. With large bandwidths, and
integration times of 35 minutes, the nominal sensitivity of these
systems is about 7 times better than their predecessors
\cite{clj+92,jlm+92}; they are thus expected to discover several
hundred new pulsars. Indeed, the Parkes multibeam survey has got off
to a flying start, discovering over 60 pulsars in the first 5\% of the
survey \cite{clb+97}.

\section{Short-Period Binary Pulsars}
\label{sec:binaries}

As pointed out by a number of authors \cite{knr90,jk91,and92},
searches with longer integration times begin to lose sensitivity to
pulsars in tight binary systems relative to a solitary pulsar. In a
short-period binary system, the apparent pulse frequency may become
significantly Doppler shifted during the integration. Thus, rather
than each harmonic being a single spike in the power spectrum, the
power may be deposited into a number of spectral bins resulting in a
net reduction in SNR.

At the cost of additional computing time, it is possible to counter
this effect by quadratically stretching or compressing the
de-dispersed time series for a number of constant trial accelerations
before rebinning and FFTing the data \cite{mp86}, In radio pulsar
searches, this technique has to date mainly been used in globular
cluster experiments where integration times may be as large as 1 hour
or more \cite{and92}. Acceleration searches have so far not been
employed in any untargetted survey with integration times $\lapp 10$
min. In such cases, good sensitivity is maintained to most binary
systems with orbital periods greater than 1--3 hr.

As mentioned in \S \ref{sec:sens}, the new multi-beam surveys have
integration times $\sim 35$ min. Here the effects of Doppler shifting
across several spectral bins can become significant.  As an example, I
generated a fake time series containing a pulsar with similar
parameters to B1913+16 for a number of initial orbital phases and fed
it through a search code with similar characteristics to the
multi-beam searches. Depending on the orbital phase, the performance
of the search code was down by typically 60\% (and in the extreme case
90\%) relative to tests on fake data for a solitary pulsar with the
same pulse period as B1913+16.

To increase the sensitivity to such binary systems (which are, after
all, the most interesting ones to detect!), an acceleration search is
currently being implemented in the multi-beam searches (A.~G.~Lyne,
private communication). Rather than a rebinning scheme outlined above,
the time series will be split into a number of segments, across which
the drift due to binary motion is reduced {\it and} the corresponding
width of frequency bins in the Fourier domain is increased. By
shifting and stacking these segments for a number of trial shifts, it
is possible to recover much of the sensitivity to short-period binary
pulsars with only a modest amount of additional computing cost.

\section{Sub-millisecond Pulsars}
\label{sec:submsps}

PSR B1937+21 still holds the record for the pulsar with the shortest
rotation period (1.56 ms), despite subsequent searches that are
theoretically sensitive to even shorter period objects.  The neutron
star equation of state is currently not well constrained. As a
consequence, the limiting spin period, below which centrifugal forces
would rip the neutron star to shreds, may vary anywhere between 0.5 ms
and just below the present period of B1937+21. Present limits on the
existence of sub-millisecond pulsars are rather poor, since signals
just above the Nyquist period of the surveys ({\it e.g.}~$\gapp 0.6$
ms for the Parkes Southern Sky survey) are significantly affected by
dispersion broadening across individual filterbank channels, with only
the very nearby (DM $\lapp 10$ cm$^{-3}$ pc) bright pulsars being
theoretically detectable. Practically, the difficulty of fishing out
such candidates, which look essentially like sinusoidal sources of
interference, means that the {\it effective} sensitivity to
sub-millisecond pulsars is presently negligible \cite{bd97}.

The future prospects of finding such pulsars look set to improve,
however, through new surveys with ever faster sampling rates and
narrower filterbank channels \cite{bd97}. In a novel 430 MHz survey,
Bailes et al.~(private communication) are presently using an S2 system
to record baseband data. Subsequent off-line processing has been
developed to simulate a filterbank in software! The present
configuration samples $1024 \times 15.625$ kHz channels every
64$\mu$s, giving much better sensitivity to sub-millisecond pulsars
than the Southern Sky Survey. With continual improvements
computing power, it seems likely that the existence of such exotic
objects will be settled in the coming years --- hopefully pinning down
our knowledge of the neutron star equation of state.

\acknowledgements{I would like to thank a number of people with whom I
have had the good fortune to hunt pulsars with and learn from over the
years {\it viz:} Andrew Lyne, Dick Manchester, Matthew Bailes, Simon
Johnston, Paul Harrison, Luciano Nicastro, Nichi D'Amico, Jon Bell,
Fernando Camilo, Vicky Kaspi and Michael Kramer. Thanks also to
Dipankar Bhattacharya for providing Fig.~\ref{fig:hsum}.}

\vfill 
\end{document}